# 'CodeAliker' - Plagiarism Detection on the Cloud

Nitish Upreti[1] and Rishi Kumar[2]

Department of Computer Science and Engineering, AMITY University, Noida, India.
nitishupreti@gmail.com

Department of Computer Science and Engineering, AMITY University, Noida, India.
rishikumar182000@gmail.com

## *Abstract*

*Plagiarism is a burning problem that academics have been facing in all of the varied levels of the educational system. With the advent of digital content, the challenge to ensure the integrity of academic work has been amplified. This paper discusses on defining a precise definition of plagiarized computer code, various solutions available for detecting plagiarism and building a cloud platform for plagiarism disclosure.*

*'CodeAliker', our application thus developed automates the submission of assignments and the review process associated for essay text as well as computer code. It has been made available under the GNU's General Public License as a Free and Open Source Software.*

## 1.Introduction

An insightful look into the scenario of academic integrity and its implications give us the major motivation for pursuing the subject. The issue holds utmost significance as the intellectual standards of an individual pursuing an academia a reestablished around his ability to produce authoritative work. Plagiarism is thus lethal. Every year a large number of students and scholars submit a huge volume of material to their respective mentors and professors. Due to the sheer amount of text involved, a manual scrutiny is infeasible. Analyzing the situation, we found no existing work in the public domain that solved the problem faced by educational institutes worldwide. Most of the alternatives were either closed source or catered to only a fraction of the entire problem. Working on this issue, at the outset we explore the sensitive aspect of classification of documents as 'authentic' or 'plagiarized'. We then analyze numerous approaches to Plagiarism detection. Advancing then to our chief goal of implementing an engine and leveraging the cloud platform for scalable and robust plagiarism detection. Alex Aliken's MOSS[1] is chosen as the key approach for building the application. Result and conclusion follow where we present our observations and learning.

## 2. Classification of Text

Broadly categorizing, the nature of text submitted to such a system can either be an Essay that is plain text in language or computer code in any of the popular language such as C, C++, Java or Ruby for instance. It is easy to figure out whether an essay text has been plagiarized however source code copying is a delicate issue with mostly a fine line drawn between 'code reuse',

'collaboration', 'non-citation' and 'plagiarism'. With learning themes such as 'Code Reuse' in an OOP system and 'Don't re-invent the wheel' code philosophies, the distinction are even more blurred. With hardly any definition in place designing a system capable of accurately identifying copying instances is infeasible. Hence concrete definitions need to be in place. A little work has been done on the topic; the only concrete input comes from the work of Georgina Cosma and Mike Joy [2]. Their work follows a survey-based approach in the U.K academics for finding the right answers and opinions. However for implementing a practical solution we need to have our own precise judgment on the problem rather than a crude hypothesis. Code assignment submitted consists of comments and the actual source code. Comments, which occur in almost all the sophisticated code bases, could be a major potential resource for identifying plagiarism instances. Nonetheless they present a major pitfall and could lead on to suspicious looking false positives as Copyright statements occur frequently as comments. For the purpose of CodeAliker we choose to strip off comments so as to avoid any such issues. Moving forward, we address a multitude of other questions. Is using an external library or API an instance of plagiarism? For most of the cases we found that library use without citation is legitimate as they are a central part of any sophisticated piece of computer program. CodeAliker thus filters out import statements and library includes. An intuitive User Interface design can also be a part of submission; code for the design is put under scrutiny by CodeAliker but looking at the design manually is much more effective than plain UI code checking.

## 3. Approaches to Plagiarism Detection

Various different approaches to Plagiarism detection exist and their performance and speed vary to a great extent. Also certain plagiarism detection schemes are more suitable for data set with a specific structure and nature. A rich taxonomy can be summarized in the diagram below.

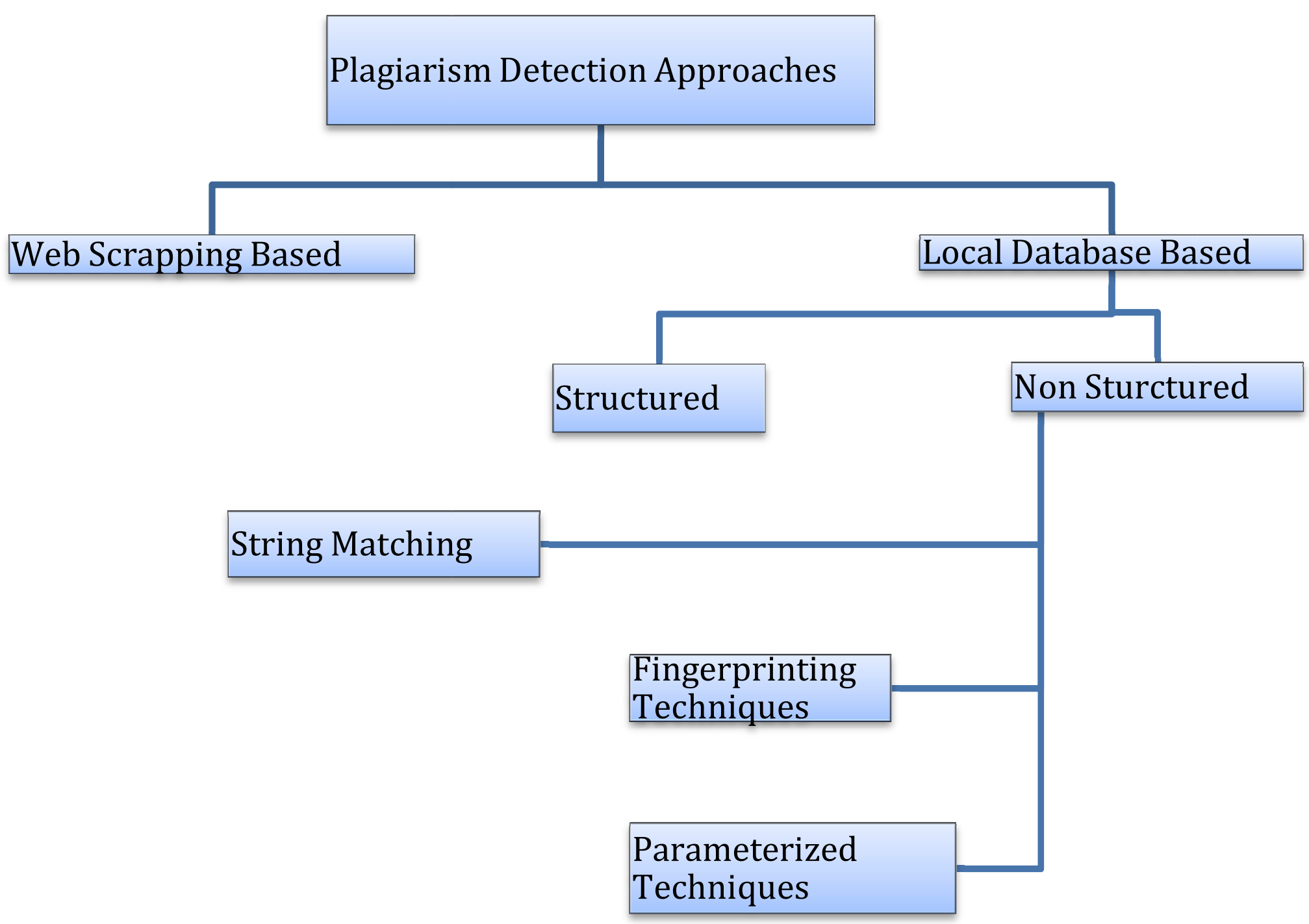

Figure 3.1

Web Scrapping based approaches use the World Wide Web to check for Plagiarism instances from a large corpus of data. The scope of Web Scrapping is huge and lots of published work exists on such systems. Our focus for this research is on systems based on a local database compiled from assignments submitted by students taking the classes and past year submissions. Local Database Based Approaches can be either Structured or Non Structured. The Structured approach creates a graph model of information in the document. This approach is used mostly with code-based assignments.

Non-Structured techniques are the most popular ones and are useful on a wide variety of text material. They are classified based on the algorithm used. Document Fingerprinting, String Matching and Parameterized Matching are the popular ones [3].

Tools based on the fingerprint approach work by creating "fingerprints" for each file which consist statistical information about the file, such as average number of terms per line, number of unique terms, and number of keywords [4].The DUP tool [5] is based on a parameterized matching algorithm, which detects identical and near-duplicate sections of source-code, by matching source-code sections whose identifiers have been substituted (renamed) systematically [3].

String Matching algorithms are quite popular and effective. MOSS [1], (YAP3) [6], JPlag [7], and Sherlock [8] are some of the popular ones available. CodeAliker is based on MOSS[1] that employs string-matching algorithms using k-grams, where a k-gram is an adjacent substring of length k. Winnowing, a local fingerprinting algorithm is also used to ensure matches of certain length are detected.

## 4. Designing the Engine with Ruby

There were various motives for choosing MOSS as the core for CodeAliker's engine. Also Ruby was used to implement the engine after considering several important factors. The language provides excellent text processing libraries, encourages an agile development methodology and Test Driven Development (TDD). Moreover it is ready for the web with excellent frameworks available.

MOSS is highly effective for plagiarism detection with text of different nature. It can also be scaled to handle a large volume of data. MOSS also guarantees matches of certain length to be detected [1].

The engine consists of three major modules: Text Filter, Hasher and Winnower. All of the components can be customized with easy to write configuration files.

The text filter has a key role to play when processing code assignments. Based on the approach MOSS suggests, the comments are stripped off, text is lowercased, identifiers are replaced with a dummy symbol, language specific keywords are removed and punctuations with no semantic meanings are stripped off. Filtered text with noise eliminated is thus obtained.

The filtered text is then fed to a Hasher that calculates hashes for the given text. A rolling hash function based on the famous Rabin Karp Algorithm is employed to calculate hashes quickly. With each hash value calculated, the corresponding line number where the text occurred is stored. This aids later in presenting user with the information regarding the instances where plagiarized text is present.

The Winnower is an implementation of the 'Robust Winnowing' algorithm defined by MOSS.A set of hash is chosen to be as the finger print of a document. Line number information is still preserved.

Winnower needs to be configured with parameters value ‘k’ for k-gram, a threshold value ‘t’ and a modulus value ‘q’. If there is a substring match at least as long as the guarantee threshold, ‘t’, then this match is detected, and we do not detect any matches shorter than the noise threshold, ‘k’ [1].The hash values computed are two large and hinder a scalable implementation; hence a value ‘q’ is used as the modulus.

For CodeAliker we found the sweet spot with the values 5(k), 8(t) and 10001(q) respectively.

The documents are compared based on the final fingerprints, with plagiarism instance being reported line by line. Check for essay based assignment is surprisingly similar with the Filter step being omitted.

## 5. Building a Cloud Application

The most interesting part of our research is to build a cloud application for the engine. For building the web application we employ the Ruby on Rails platform.

Ruby on Rails, a full stack framework for Ruby is excellent for agile development and sustainable productivity. It boasts a high modular design, excellent package management capabilities, database abstraction with ORM(Object Relational Mapping) library and ease of deployment.

The application is built with the MVC (Model – View – Controller) design pattern inherent on the Rails Framework. CodeAliker aims to ease the workflow involved by automating the entire process. To achieve this, an authentication-based system is introduced for the professors where assignments for each class they take are available to them as a separate bunch.

The professor can mark any on the assignment as primary and check with respect to that assignment all the possible plagiarized instances. Thus getting a complete view of the scenario effortlessly in a non ad-hoc fashion. Academics can also manually supervise the submission and reviews.

While the traditional delivery of software services have been mainstream, bringing the cloud into perspective changes the entire scenario. Cloud tends to centralize our resources, code base and data onto an always-available depot. Hardware resources can be accurately utilized, load on a high demand system can be catered to and system can be easily scaled. Configuration management is also made effortless. Any organization with requirements for sucha system is in need of the cloud. The platform allows changes to be pushed onto codebase with a push of a button, rather than relying on extensive upgrade packs.

For plagiarism detection in a university or an academic institute the needs are critical and point towards the cloud. The computing requirements are thus addressed. Also our system is centralized and easily scaled.

For hosting CodeAliker on the cloud, Heroku, a cloud platform has been employed.
The source code is available to public here on: https://github.com/Myth17/CodeAliker
The application is available for free use at: http://codealiker.heroku.com/

# 6. Results

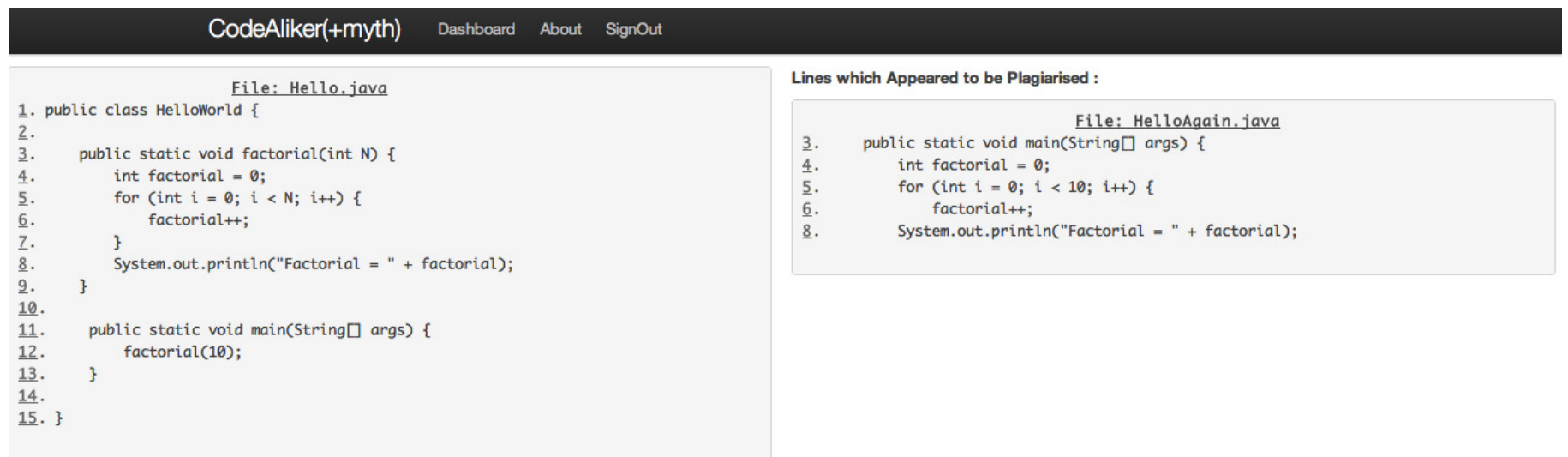

Figure 6.1

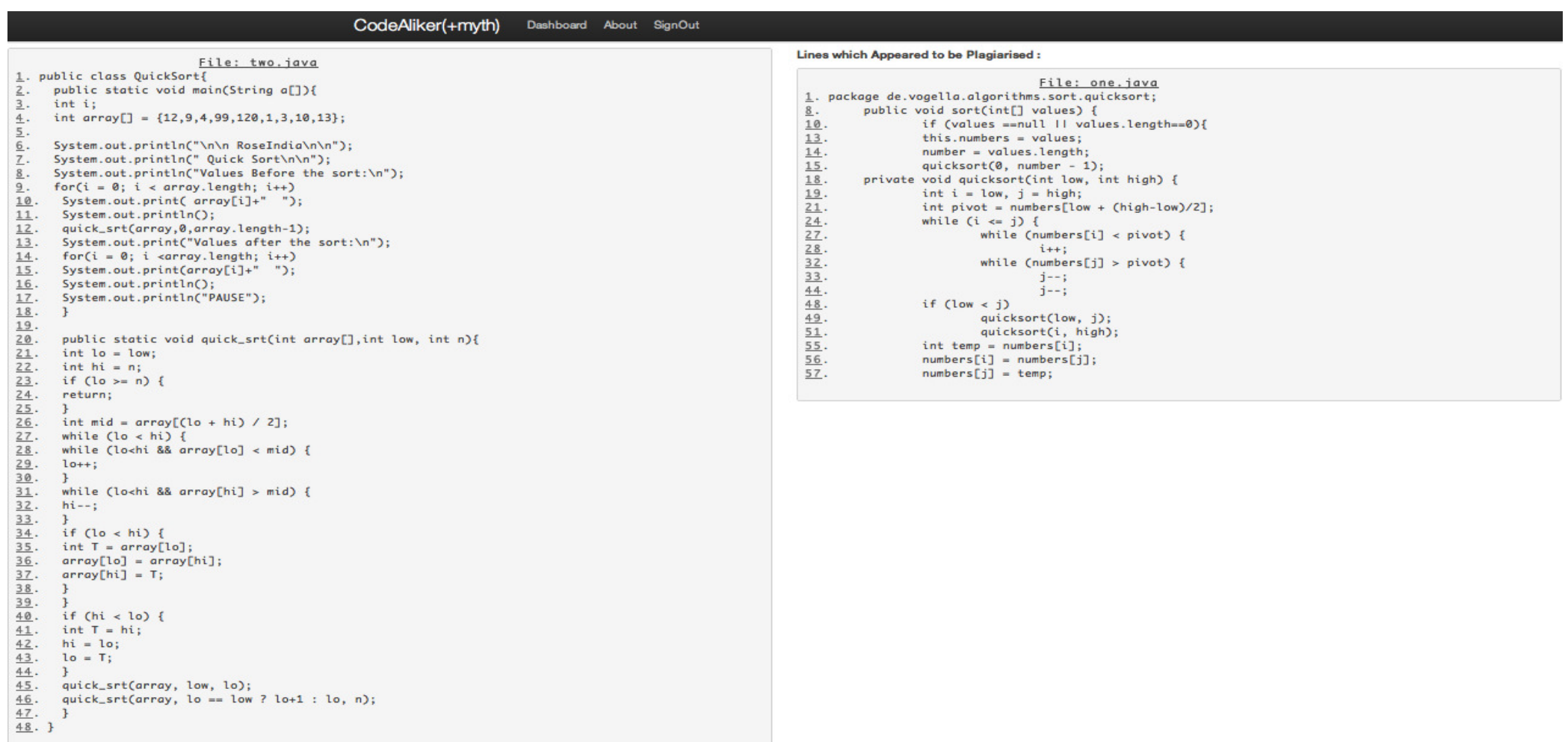

Figure 6.2

The results for CodeAliker display the assignment marked as primary to the left while the suspected plagiarism instances are previewed stacked onto each other in the right. To present a clearer picture, the plagiarized instances are marked with the line numbers in order of aid manual scrutiny and presenting a more cohesive report.

# 7. Conclusion

We have analyzed the entire scenario of Plagiarism detection, while figuring out the problems and solutions for developing an application for the purpose. Major accomplishment of the research being our ability to define a precise definition of code plagiarism, understanding different approaches towards plagiarism detection, practical implementation of the MOSS engine with fine tuned parameters and building a scalable web application hosted on a cloud platform.

### Authors

**Nitish Upreti** is computer science student at AMITY School of Engineering and Technology, Noida, India. His fields of interest include Algorithm Design, Artificial Intelligence and building scalable web applications.

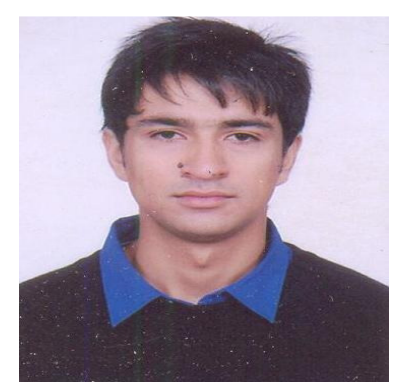

**Rishi Kumar** is computer science faculty at Amity School of Engineering and Technology, Noida, India. His fields of interest include Artificial Intelligence, Expert System & Image Processing.

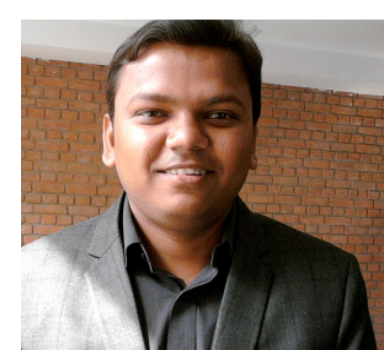